\documentclass[aps,showpacs,preprintnumbers,floatfix]{revtex4}
\usepackage{graphicx}
\usepackage{epsfig}
\usepackage{dcolumn}
\usepackage{subfigure}
\usepackage{amsmath}

\def\be{\begin{equation}}
\def\ee{\end{equation}}
\def\bea{\begin{eqnarray}}
\def\eea{\end{eqnarray}}

\newcommand{\omits}[1]{}

\def\bsp{\be\begin{split}}

\def\bes{\be  \begin{split}}

\newcommand{\Rmnum}[1]{\expandafter\@slowromancap\romannumeral #1@}

\def\NPB{{Nucl. Phys.}~{\bf B}}
\def\PRD{{Phys. Rev.}~{\bf D}}

\def\JHEP{{JHEP}}

\begin{document}

\title{Phase transition of the higher dimensional
charged Gauss-Bonnet black hole in de Sitter spacetime}

\author{Meng-Sen Ma$^{a,b}$\footnote{Email:mengsenma@gmail.com}, Li-Chun Zhang$^{a,b}$\footnote{Email:zhao2969@sina.com}, Hui-Hua Zhao$^{a,b}$\footnote{Email:kietemap@126.com}, Ren Zhao$^{a,b}$\footnote{Email:zhao2969@sina.com}}

\medskip

\affiliation{\footnotesize$^a$Department of Physics, Shanxi Datong
University,  Datong 037009, China\\
\footnotesize$^b$Institute of Theoretical Physics, Shanxi Datong
University, Datong 037009, China}

\begin{abstract}

We study the phase transition of charged Gauss-Bonnet-de Sitter (GB-dS) black hole. For black holes in de Sitter spacetime, there is not only  black hole horizon, but also the cosmological horizon. The thermodynamic
quantities on the both horizons satisfy the first law
of the black hole thermodynamics, respectively; moreover,
there are additional connections between them. Using the effective temperature approach, we obtained the effective thermodynamic quantities of charged GB-dS black hole.
According to Ehrenfest classification, we calculate some response functions
and plot their figures, from which one can see that the spacetime undergoes a second-order phase transition
at the critical point. It is shown that the critical values of effective temperature and pressure decrease with the increase of the value of GB parameter $\alpha$.

\end{abstract}

 \maketitle

\section{Introduction}

Due to the existence of Hawking radiation and the entropy, black holes may display themselves like thermodynamic systems\cite{Bekenstein,Hawking1,Hawking2,Hawking3}.
Phase transitions and critical phenomena are important characteristics of ordinary
thermodynamic system. Thus, the natural question to ask is whether there also exists phase transition
in the black hole thermodynamics. The pioneer work of Hawking and Page give us an definitive answer\cite{Hawking4}.
Recently, the phase transitions of black holes in asymptotically anti de-Sitter
(AdS) spacetime have received considerable attention\cite{Chamblin1,Chamblin2,Sarkar1,Sarkar2,Sarkar3,Banerjee1,Banerjee2,Banerjee3,Banerjee4,Banerjee5}. In particular, by considering the cosmological constant as thermodynamic pressure
$P=-\frac{\Lambda}{8\pi } =\frac{(d-1)(d-2)}{16\pi l^2}$, one can introduce an extended phase space,
in which the $P-V$ criticality has been extensively discussed\cite{Kastor,Dolan1,SG,Dolan2,Cvetic,Mann2,Dolan3,LYX,Cai1,Liu,Hendi,Zhao1,Mann3,LYX2,Dolan4,BZ}.
It is shown that many black holes in AdS spacetime exhibit similar critical behaviors to that of Van der Waals liquid-gas system.

The astronomical observations show that our Universe is
probably an asymptotically de Sitter (dS) one. It
raises the interest on black holes in dS spacetime.
There are many works on the thermal properties of black holes in de Sitter(dS) spacetime\cite{Cai2,Sekiwa,Urano,Zhao2,Dolan4,Myung,Cai3,SB,Ma,Cai4,Gibbons1,Gibbons2}.
As is well known, there are multi-horizons for black holes in dS spacetime.
 On the black hole event horizon and the cosmological horizon there are different temperatures, which prevent the black hole thermodynamic system in equilibrium.
It should be noted that there are common parameters $M$, $Q$, $\Lambda$ in the thermodynamic quantities corresponding to the black hole horizon and that corresponding to the cosmological horizon. Thus, these thermodynamic quantities are not independent. Taking into account their connections is relevant to the understanding of thermal properties of dS spacetime.

Higher derivative curvature terms occur in many occasions,
such as in the semiclassically quantum gravity and in
the effective low-energy action of superstring theories.
Among the many theories of gravity with higher derivative curvature
terms, due to the special features the Gauss-Bonnet (GB) gravity has attract much interest. The thermodynamic properties and phase structures of GB-AdS black hole
have been briefly discussed in \cite{Cai5}. In Refs.\cite{LYX,Cai1,Liu}, the critical phenomena and phase transition of the charged GB-AdS black hole have been studied extensively.

In this paper, we study the thermal properties of charged GB-dS black hole after considering the connections between the black hole horizon and the cosmological horizon.
Using the effective equilibrium temperature approach, we calculate some effective thermodynamic quantities, from which we can analyze whether there exists phase transition for
charged GB-dS black hole.

The paper is arranged as follows: in the next section we simply
introduce the charged Gauss-Bonnet-dS black hole. In section 3,  we will
calculate the effective thermodynamic quantities. In section 4, we derive the heat capacity at constant pressure,  the
volume expansivity $\beta$, and the isothermal compressibility $\kappa$ and plot some curves to demonstrate the phase transition. We will make some
concluding remarks in section 5.
(we use the units $G_d =\hbar =c=k_B =1)$

\section{Charged Gauss-Bonnet-dS black hole}

The action of the $d$- dimensional Einstein-Gauss-Bonnet-Maxwell-dS theory
has the form
\begin{equation}
\label{eq2}
S=\frac{1}{16\pi }\int {d^d} x\sqrt {-g} [R-2\Lambda +\alpha (R_{\mu \nu
\gamma \delta } R^{\mu \nu \gamma \delta }-4R_{\mu \nu } R^{\mu \nu
}+R^2)-4\pi F_{\mu \nu } F^{\mu \nu }],
\end{equation}
where $\alpha $ is the Gauss-Bonnet coupling constant and the cosmological
constant is $\Lambda =\frac{(d-1)(d-2)}{2l^2}$ for dS spacetime, $F_{\mu \nu
} $ is the Maxwell field strength. The $d$-dimensional static charged
Gauss-Bonnet-dS black hole solution for the action is described by
\begin{equation}
\label{eq3}
ds^2=-fdt^2+f^{-1}dr^2+r^2(d\theta ^2+\sin ^2d\phi ^2+\cos ^2\theta d\Omega
_{d-4}^2 ),
\end{equation}
with the metric function given by\cite{LYX,Cai1,LYX2}
\begin{equation}
\label{eq4}
f(r)=1+\frac{r^2}{2\tilde {\alpha }}\left( {1-\sqrt {1+\frac{64\pi \tilde
{\alpha }M}{(d-2)\Sigma _k r^{d-1}}-\frac{2\tilde {\alpha
}Q^2}{(d-2)(d-3)r^{2d-4}}+\frac{8\tilde {\alpha }\Lambda }{(d-1)(d-2)}} }
\right),
\end{equation}
where $\tilde {\alpha }=(d-3)(d-4)\alpha $, $\Sigma _k $ is the area of a
unite $(d-2)$-dimensional sphere, $M$ is the black hole mass, $Q$ is related
to charge of the black hole. Horizons occur where $f(r)=0$, the largest root
is the cosmological horizon $r=r_c $, and the next root is the black hole
event horizon $r=r_+ $, When $M=Q=0$, the solution (\ref{eq3}) reduces to the pure
GB-dS space with only one cosmological horizon at $r_c =l_c $.

The equations $f(r_+ )=0$ and $f(r_c )=0$ are rearranged to
\[
M=\frac{(d-2)\Sigma _k r_+^{d-3} }{16\pi }\left( {1+\frac{\tilde {\alpha
}}{r_+^2 }} \right)-\frac{\Sigma _k r_+^{d-1} \Lambda }{8\pi
(d-1)}+\frac{\Sigma _k Q^2}{32\pi (d-3)r_+^{d-3} },
\]
\[
M=\frac{(d-2)\Sigma _k r_c^{d-3} }{16\pi }\left( {1+\frac{\tilde {\alpha
}}{r_c^2 }} \right)-\frac{\Sigma _k r_c^{d-1} \Lambda }{8\pi
(d-1)}+\frac{\Sigma _k Q^2}{32\pi (d-3)r_c^{d-3} },
\]
from which one can derive
\begin{equation}
\label{eq5}
M=\frac{(d-2)\Sigma _k r_+^{d-3} r_c^{d-3} (r_c^2 -r_+^2 )}{16\pi (r_c^{d-1}
-r_+^{d-1} )}\left( {1+\frac{\tilde {\alpha }(r_c^2 +r_+^2 )}{r_c^2 r_+^2 }}
\right)+\frac{\Sigma _k Q^2(r_c^{2d-4} -r_+^{2d-4} )}{32\pi (d-3)r_c^{d-3}
r_+^{d-3} (r_c^{d-1} -r_+^{d-1} )},
\end{equation}
by eliminating the $\Lambda $, and
\[
\Lambda =\frac{(d-1)(d-2)}{2(r_c^{d-1} -r_+^{d-1} )}\left( {r_c^{d-3}
-r_+^{d-3} +\tilde {\alpha }(r_c^{d-5} -r_+^{d-5} )}
\right)-\frac{(d-1)Q^2(r_c^{d-3} -r_+^{d-3} )}{4(d-3)r_c^{d-3} r_+^{d-3}
(r_c^{d-1} -r_+^{d-1} )}
\]
by eliminating $M$ .

The surface gravities of black hole horizon and the cosmological horizon are
\[
\kappa _+ =\left. {\frac{f'(r)}{2}} \right|_{r=r_+ }
=-\frac{1}{r_+ }+\frac{(d-1)r_c^{d-3} r_+ (r_c^2 -r_+^2 )}{2(r_+^2 +2\tilde
{\alpha })(r_c^{d-1} -r_+^{d-1} )}\left( {1+\frac{\tilde {\alpha }(r_c^2
+r_+^2 )}{r_c^2 r_+^2 }} \right)
\]
\begin{equation}
\label{eq6}
-\frac{Q^2}{4(d-3)(r_+^2 +2\tilde {\alpha })r_+^{2d-7} }\left(
{2-\frac{(d-1)(r_c^{2d-4} -r_+^{2d-4} )}{(d-2)r_c^{d-3} (r_c^{d-1}
-r_+^{d-1} )}} \right),
\end{equation}
\[
\kappa _c =\left. {\frac{f'(r)}{2}} \right|_{r=r_c }
=-\frac{1}{r_c }+\frac{(d-1)r_+^{d-3} r_c (r_c^2 -r_+^2 )}{2(r_c^2 +2\tilde
{\alpha })(r_c^{d-1} -r_+^{d-1} )}\left( {1+\frac{\tilde {\alpha }(r_c^2
+r_+^2 )}{r_c^2 r_+^2 }} \right)
\]
\begin{equation}
\label{eq7}
-\frac{Q^2}{4(d-3)(r_c^2 +2\tilde {\alpha })r_c^{2d-7} }\left(
{2-\frac{(d-1)(r_c^{2d-4} -r_+^{2d-4} )}{(d-2)r_+^{d-3} (r_c^{d-1}
-r_+^{d-1} )}} \right).
\end{equation}
The thermodynamic quantities for the two horizons satisfy the first law of
black hole thermodyamic\cite{Dolan4,Sekiwa}
\begin{equation}
\label{eq8}
\delta M=\frac{\kappa _+ }{2\pi }\delta S_+ +\Phi _+ \delta Q+V_+ \delta
P+\tilde {V}_+ d\tilde {\alpha },
\end{equation}
\begin{equation}
\label{eq9}
\delta M=\frac{\kappa _c }{2\pi }\delta S_c +\Phi _c \delta Q+V_c \delta
P+\tilde {V}_c d\tilde {\alpha },
\end{equation}
where
\[
S_+ =\frac{\Sigma _k r_+^{(d-2)} }{4}\left( {1+\frac{2(d-2)\tilde {\alpha
}}{(d-4)r_+^2 }} \right),
\quad
S_c =\frac{\Sigma _k r_c^{(d-2)} }{4}\left( {1+\frac{2(d-2)\tilde {\alpha
}}{(d-4)r_c^2 }} \right),
\]
\[
\Phi _+ =\frac{\Sigma _k Q}{16\pi (d-3)r_+^{d-3} },
\Phi _c =\frac{\Sigma _k Q}{16\pi (d-3)r_c^{d-3} },
\quad
V_+ =\frac{\Sigma _k r_+^{d-1} }{d-1},
\quad
V_c =\frac{\Sigma _k r_c^{d-1} }{d-1},
\]
\begin{equation}
\label{eq10}
\tilde {V}_+ =\frac{(d-2)}{16\pi }r_+^{d-5} ,
\quad
\tilde {V}_c =\frac{(d-2)}{16\pi }r_c^{d-5} ,
\quad
P=-\frac{\Lambda }{8\pi }.
\end{equation}

\section{effective thermodynamic quantities of charged Gauss-Bonnet black hole}

We have given the thermodynamic quantities without considering
the relationship between the black hole horizon and the cosmological
horizon. Because there are four variables $M$, $Q$ , $\Lambda$ and $\tilde{\alpha}
$ in the spacetime, the thermodynamic quantities corresponding to the black
hole horizon and the cosmological horizon are functions of these variables.
Through the four variables, there are some connections between the thermodynamic quantities corresponding to
the black hole horizon and the ones corresponding to the
cosmological horizon. When the thermodynamic property of charged de Sitter
spacetime is studied, we must consider the relationship with the two
horizon. Recently, by studying Hawking radiation of de Sitter spacetime,
refs, \cite{Zhao3,Zhao4,Zhao5} obtained that the outgoing rate of the charged de Sitter
spacetime which radiates particles with energy $\omega $ is
\begin{equation}
\label{eq11}
\Gamma =e^{\Delta S_+ +\Delta S_c},
\end{equation}
where $\Delta S_+ $ and $\Delta S_c $ are Bekenstein-Hawking entropy
difference corresponding to the black hole horizon and the cosmological
horizon after the charged de Sitter spacetime radiates particles with energy
$\omega $. Therefore, the thermodynamic entropy of the charged de Sitter
spacetime is the sum of the black hole horizon entropy and the cosmological
horizon entropy:
\begin{equation}
\label{eq12}
S=S_+ +S_c .
\end{equation}
Recently, thermodynamic volume of charged dS black hole is given by\cite{Dolan4, Sekiwa}
\begin{equation}
\label{eq13}
V=V_c -V_+ .
\end{equation}
From (\ref{eq9}), one can obtain
\begin{equation}
\label{eq14}
dS_+ =\frac{\Sigma _k (d-2)}{4}r_+^{d-5} \left( {r_+^2 +2\tilde {\alpha }}
\right)dr_+ ,
\quad
dS_c =\frac{\Sigma _k (d-2)}{4}r_c^{d-5} \left( {r_c^2 +2\tilde {\alpha }}
\right)dr_c .
\end{equation}
Substituting (\ref{eq12}) and (\ref{eq13}) into (\ref{eq14}), we obtain
\[
dS_+ =\frac{r_c^3 (r_+^2 +2\tilde {\alpha })dS}{r_c^3 (r_+^2 +2\tilde
{\alpha })+r_+^3 (r_c^2 +2\tilde {\alpha })}-\frac{(d-2)(r_+^2 +2\tilde
{\alpha })(r_c^2 +2\tilde {\alpha })dV}{4[r_c^3 (r_+^2 +2\tilde {\alpha
})+r_+^3 (r_c^2 +2\tilde {\alpha })]},
\]
\begin{equation}
\label{eq15}
dS_c =\frac{r_+^3 (r_c^2 +2\tilde {\alpha })dS}{r_c^3 (r_+^2 +2\tilde
{\alpha })+r_+^3 (r_c^2 +2\tilde {\alpha })}+\frac{(d-2)(r_+^2 +2\tilde
{\alpha })(r_c^2 +2\tilde {\alpha })dV}{4[r_c^3 (r_+^2 +2\tilde {\alpha
})+r_+^3 (r_c^2 +2\tilde {\alpha })]}.
\end{equation}
Substituting (\ref{eq15}) into (\ref{eq8}) and (\ref{eq9}), we get
\begin{equation}
\label{eq16}
dM=T_{eff} dS-P_{eff} dV+\Phi _{eff} dQ+\tilde {V}_{eff} d\tilde {\alpha },
\end{equation}
where the effective temperature $T_{eff} $, effective electric potential
$\Phi _{eff} $ and effective pressure $P_{eff} $ are respectively
\[
T_{eff} =\frac{\kappa _+ V_c r_c^3 (r_+^2 +2\tilde {\alpha })-\kappa _c V_+
r_+^3 (r_c^2 +2\tilde {\alpha })}{2\pi (V_c -V_+ )[r_c^3 (r_+^2 +2\tilde
{\alpha })+r_+^3 (r_c^2 +2\tilde {\alpha })]},
\quad
\Phi _{eff} =\frac{\Phi _+ V_c -\Phi _c V_+ }{V_c -V_+ },
\]
\begin{equation}
\label{eq17}
P_{eff} =\frac{(d-2)(r_+^2 +2\tilde {\alpha })(r_c^2 +2\tilde {\alpha
})}{8\pi [r_c^3 (r_+^2 +2\tilde {\alpha })+r_+^3 (r_c^2 +2\tilde {\alpha
})]}\frac{(\kappa _+ V_c +\kappa _c V_+ )}{(V_c -V_+ )}.
\end{equation}
\[
\tilde {V}_{eff} =\frac{V_c \tilde {V}_+ -V_+ \tilde {V}_c }{V_c -V_+ }
\]
Substituting (\ref{eq6}), (\ref{eq7}) and (\ref{eq10}) into (\ref{eq17}), we obtain

\[
T_{eff} =\frac{x[(d-1)(1-x^2)(1+x^{d-1})-2(1-x^{d+1})]}{4\pi r_c
(1-x^{d-1})[(x^2+2\tilde {\alpha }/r_c^2 )+x^3(1+2\tilde {\alpha }/r_c^2
)]}
\]
\[
+\frac{\tilde {\alpha }[(d-1)(1-x^2)(1+x^{d-1})(1+x^2)-4(1-x^{d+3})]}{4\pi
r_c^3 x(1-x^{d-1})[(x^2+2\tilde {\alpha }/r_c^2 )+x^3(1+2\tilde {\alpha
}/r_c^2 )]}
\]
\begin{equation}
\label{eq18}
+\frac{Q^2}{8\pi
(d-3)(d-2)}\frac{(d-1)(1+x^{d-1})(1-x^{2d-4})-2(d-2)(1-x^{3d-5})}{r_c^{2d-5}
x^{2d-7}(1-x^{d-1})[(x^2+2\tilde {\alpha }/r_c^2 )+x^3(1+2\tilde {\alpha
}/r_c^2 )]},
\end{equation}

\[
P_{eff} =-\frac{(d-2)(x^2+2\tilde {\alpha }/r_c^2 )(1+2\tilde {\alpha
}/r_c^2 )(1+x^d)}{4\pi r_c^2 x(1-x^{d-1})[(x^2+2\tilde {\alpha }/r_c^2
)+x^3(1+2\tilde {\alpha }/r_c^2 )]}
\]
\[
+\frac{(d-2)(d-1)(1-x^2)x}{8\pi r_c^2 (1-x^{d-1})^2}\left(
{\frac{(1+x^{2d-3})+2\tilde {\alpha }(1+x^{2d-5})/r_c^2 }{(x^2+2\tilde
{\alpha }/r_c^2 )+x^3(1+2\tilde {\alpha }/r_c^2 )}} \right)
\]
\[
+\frac{\tilde {\alpha }(d-2)(d-1)(1-x^2)(1+x^2)}{8\pi r_c^4
(1-x^{d-1})^2x}\left( {\frac{(1+x^{2d-3})+2\tilde {\alpha
}(1+x^{2d-5})/r_c^2 }{(x^2+2\tilde {\alpha }/r_c^2 )+x^3(1+2\tilde {\alpha
}/r_c^2 )}} \right)
\]
\[
-\frac{Q^2(d-2)}{8\pi (d-3)r_c^{2d-4} x^{2d-7}(1-x^{d-1})}\left(
{\frac{(1+x^{3d-6})+2\tilde {\alpha }(1+x^{3d-8})/r_c^2 }{(x^2+2\tilde
{\alpha }/r_c^2 )+x^3(1+2\tilde {\alpha }/r_c^2 )}} \right)
\]
\begin{equation}
\label{eq19}
+\frac{Q^2(d-1)(1-x^{2d-4})}{16\pi (d-3)r_c^{2d-4}
x^{2d-7}(1-x^{d-1})^2}\left( {\frac{(1+x^{2d-3})+2\tilde {\alpha
}(1+x^{2d-5})/r_c^2 }{(x^2+2\tilde {\alpha }/r_c^2 )+x^3(1+2\tilde {\alpha
}/r_c^2 )}} \right),
\end{equation}
\begin{equation}
\label{eq20}
\Phi _{eff} =\frac{\Sigma _k Q(1-x^{2d-4})}{16\pi (d-3)r_c^{d-3} x^{d-3}(1-x^{d-1})}.
\end{equation}
where $x:=r_+ /r_c $ and $0<x<1$.

In the case of neglecting the connections between the two horizons, the
black hole horizon and the cosmological horizon are two independent
thermodynamic systems. Due to different horizon temperatures, the
spacetime cannot be in thermal equilibrium. After considering the
connections between the both horizons, from Eq.(\ref{eq18}) one can see that only
one effective temperature $T_{eff} $ left.

\section{Phase transition in Gauss-Bonnet-dS black hole spacetime}

Recently, there are many works on the phase transition of black hole thermodynamic system.
One can also compare the critical behaviors of van der Waals liquid-gas system and black hole system.
Nevertheless, in dS spacetime there are the cosmological horizon and the black hole horizon. Generally, equilibrium cannot be achieved unless for extremal black hole. In this section, we employe the effective thermodynamic quantities to study the phase transition and critical behaviors of charged Gauss-Bonnet-dS black hole. This may avoid the apprehension of equilibrium of the black hole system.

Comparing with the Van der Waals equation
\begin{equation}
\label{eq25}
\left( {P+\frac{a}{v^2}} \right)(v-\tilde {b})=kT,
\end{equation}
Here, $v=V/N$ is the specific volume of the fluid, $P$ its pressure, $T$ its
temperature, and $k$ is the Boltzmann constant. Similarly, for the GB-dS black hole we can set the specific volume as
\be
v=r_c(1-x).
\ee

According to Eq.(\ref{eq19}), one can employ the two equations
\begin{equation}
\label{eq31}
\left( {\frac{\partial P_{eff} }{\partial x}} \right)_{T_{eff} } =0,
\quad
\left( {\frac{\partial ^2P_{eff} }{\partial x^2}} \right)_{T_{eff} } =0,
\end{equation}
to calculate the critical temperature, critical pressure and the critical specific volume.
We can ascertain the critical values ($r_c^0
,x^0)$ for given $Q, n$ and $\tilde \alpha$. According to  these values,
we can obtain the critical temperature, critical electric potential, critical pressure and critical volume. As is shown Table I, the critical temperature and critical pressure increase with the dimension of spacetime for fixed $Q$ and $\tilde\alpha$. However, they decrease with the increase of the Gauss-Bonnet parameter $\tilde\alpha$ for fixed spacetime dimension and electric charge $Q$.

\begin{table}[!htbp]
\begin{center}
\begin{tabular}{|c|c|c|c|c|c|c|}
\hline
\raisebox{-1.50ex}[0cm][0cm]{}&
\multicolumn{3}{|c|}{$T_{eff}^c $} &
\multicolumn{3}{|c|}{$P_{eff}^c $}\\
\cline{2-7}
&$n=5$&
$n=6$&
$n=7$&
$n=5$&
$n=6$&
$n=7$
 \\
\hline
$\tilde \alpha=0.1$&
0.028628&
0.040562&
0.049202&
0.008229&
0.018877&
0.031920
 \\
\hline
$\tilde \alpha=0.01$&
0.033651&
0.045736&
0.054251&
0.010356&
0.021471&
0.034484 \\
\hline
$\tilde \alpha=0.001$&
0.034297&
0.046385&
0.054889&
0.010632&
0.021802&
0.034836
\\
\hline
\end{tabular}
\label{tab1}
\end{center}
\caption{Numerical solutions for $T_{eff}^c$, $P_{eff}^c$ with fixed $Q=1$ and spacetime
dimension $n=5,6,7$ and GB parameter $\tilde \alpha=0.1,0.01,0.001$, respectively.}
\end{table}

In order to describe the relation of $P_{eff} $ and $V$ in the vicinity of critical temperature, we plot the curves of $P_{eff} $-$V$ at different temperatures.
When the effective temperature $T_{eff} >T_{eff}^c $,
the stable condition $\left( {\frac{\partial P}{\partial V}}
\right)_{T_{eff} } <0$ can be satisfied. In the case of  $T_{eff}
<T_{eff}^c $, there exist a portion in the curve where
$\left({\frac{\partial P}{\partial V}} \right)_{T_{eff} }
>0$, thus in these parts the system is unstable. So phase transition may occur only at $T_{eff} =T_{eff}^c $.

\begin{figure}[htbp]
\center{\subfigure[~$ d=5 $, $\tilde \alpha=0.1$, $T_c=0.028628$] {
\includegraphics[width=4cm,keepaspectratio]{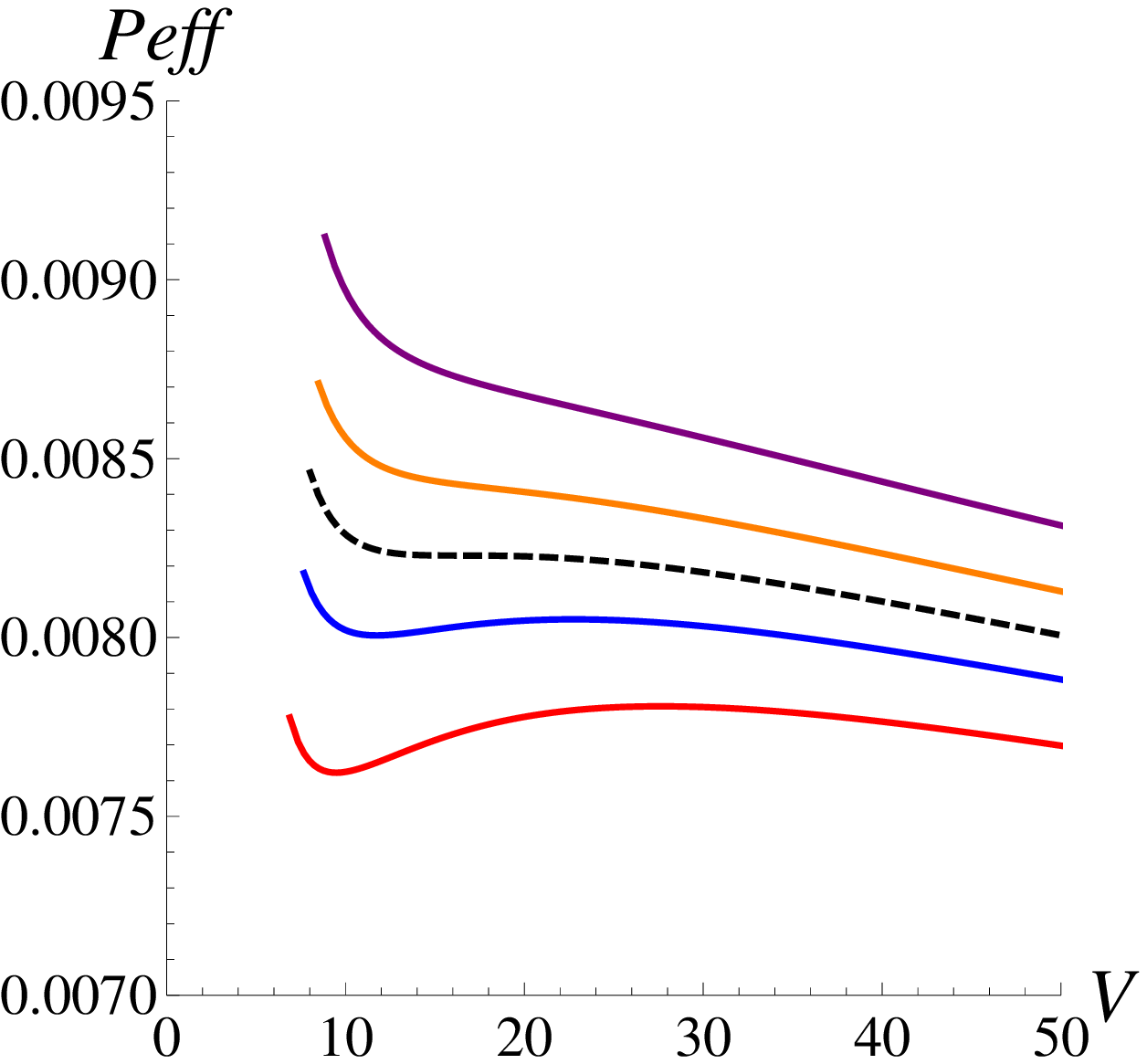}\hspace{0.8cm}}
\subfigure[~$ d=6$, $\tilde \alpha=0.1$, $T_c=0.040562$] {
\includegraphics[width=4cm,keepaspectratio]{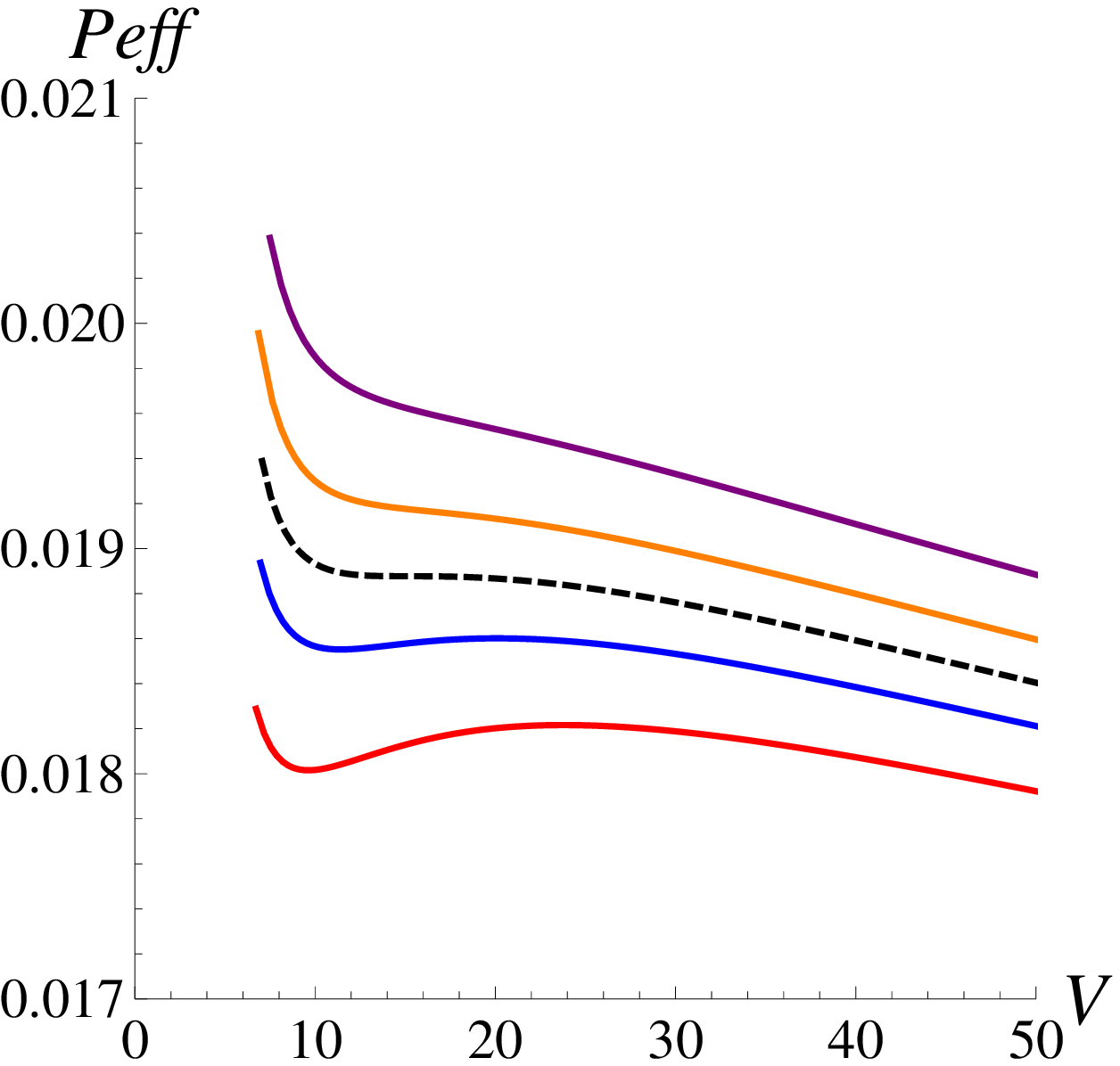}\hspace{0.8cm}}
\subfigure[~$ d=7$, $\tilde \alpha=0.1$, $T_c=0.049202$] {
\includegraphics[width=4cm,keepaspectratio]{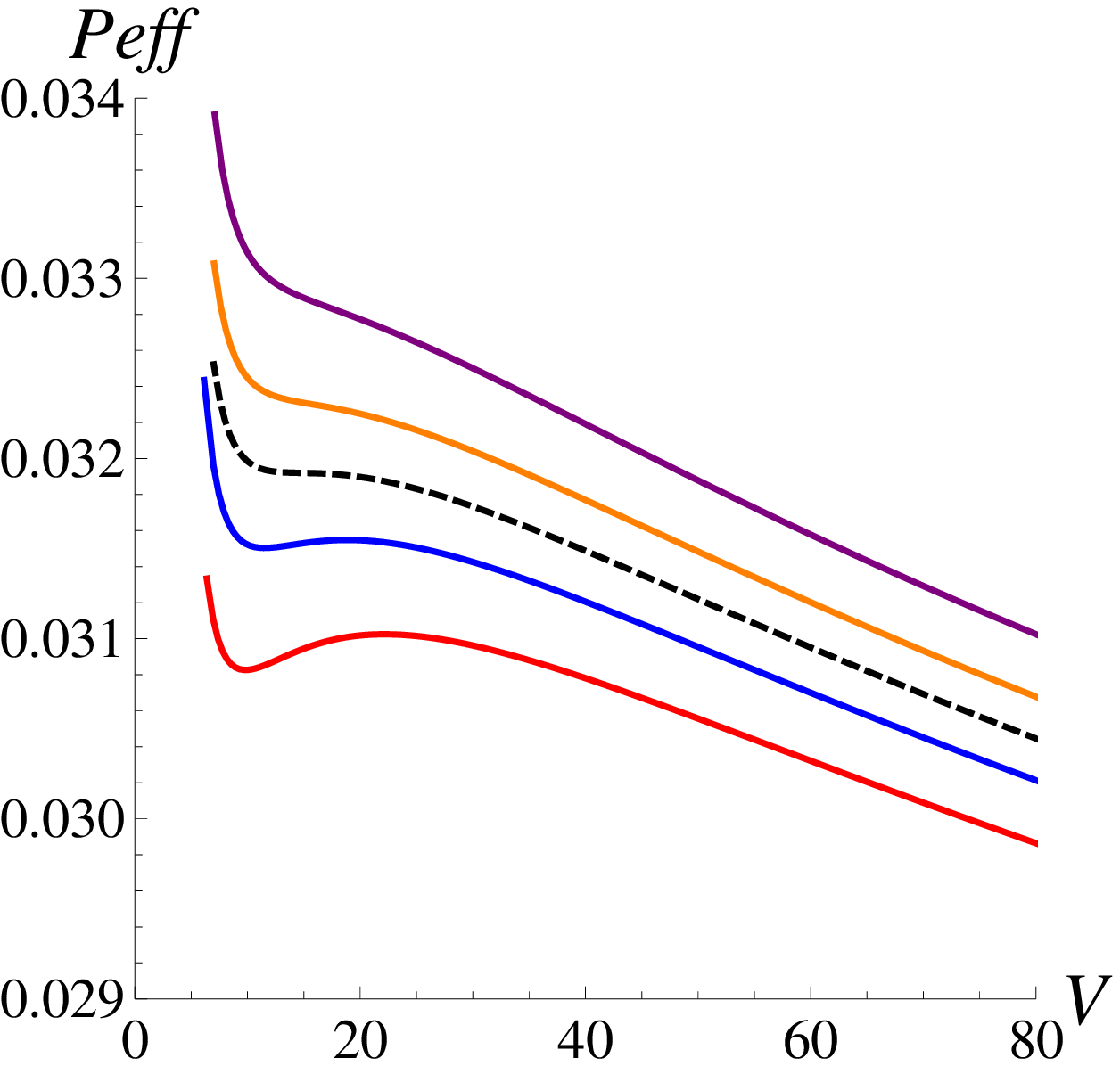}\hspace{0.8cm}}\\
\subfigure[~$ d=5 $, $\tilde \alpha=0.01$, $T_c=0.033651$] {
\includegraphics[width=4cm,keepaspectratio]{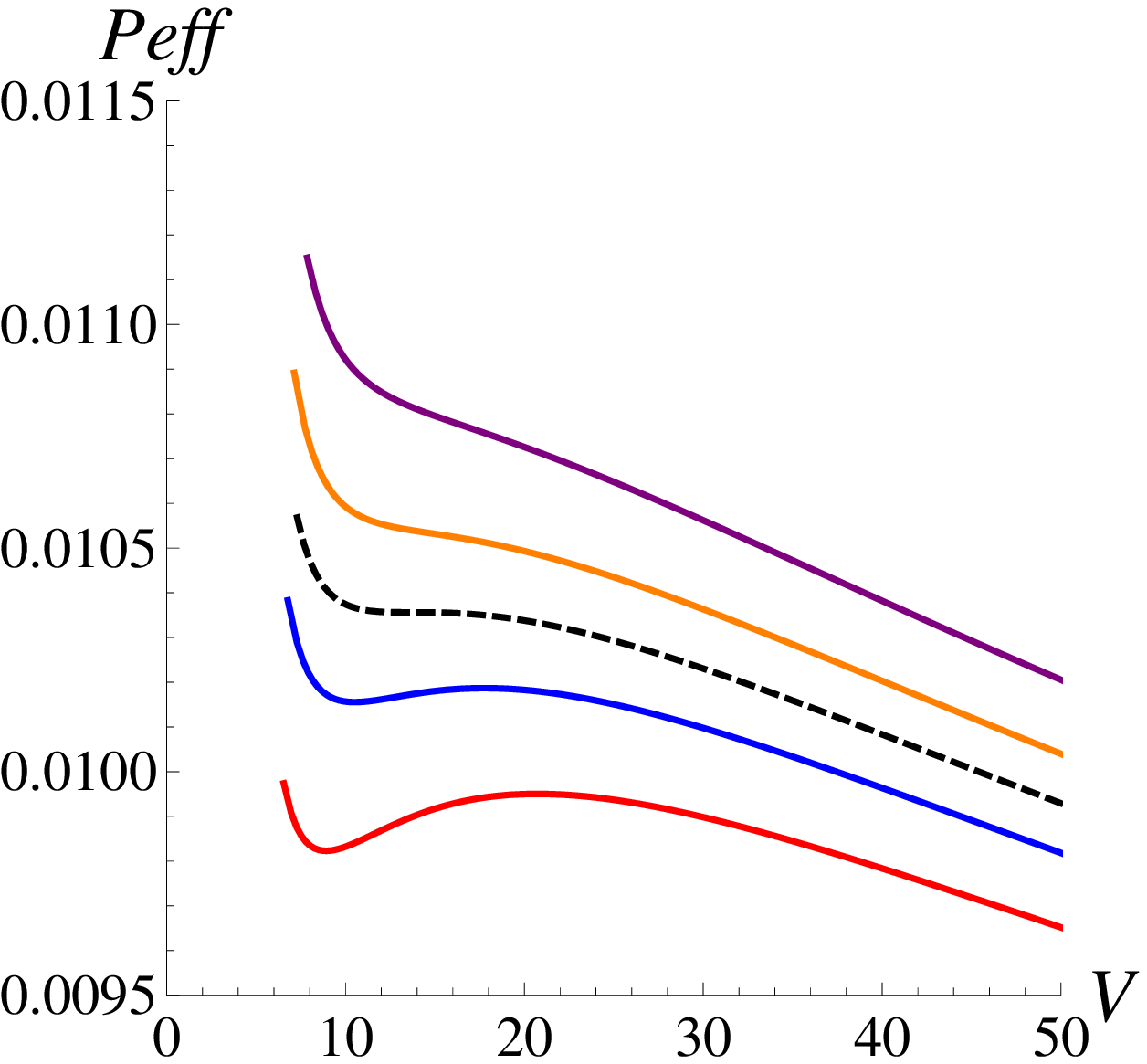}\hspace{0.8cm}}
\subfigure[~$ d=5 $, $\tilde \alpha=0.001$, $T_c=0.034297$] {
\includegraphics[width=4cm,keepaspectratio]{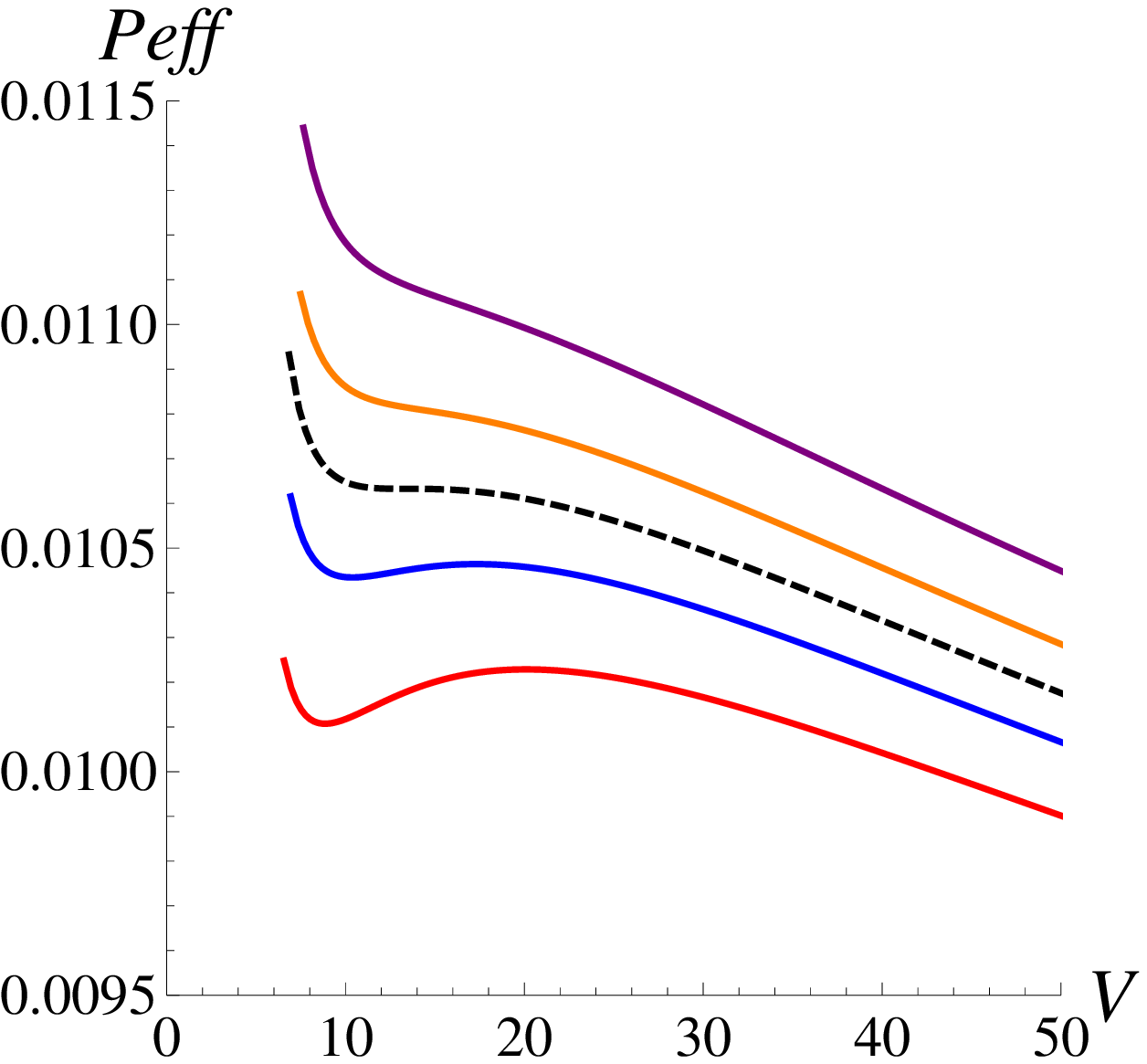}\hspace{0.8cm}}
\caption{$P_{eff}-V$ diagram of charged GB-dS black hole. From top to bottom the curves correspond to the effective
temperature $T_{eff}^c +0.0005$, $T_{eff}^c +0.0002$, $T_{eff}^c$,
$T_{eff}^c -0.0002$ and $T_{eff}^c -0.0005$. We have set $Q=1$.}}\label{fig1}
\end{figure}

According to Ehrenfest classification, when the chemical potential and its first derivative are continuous, whereas
the second derivative of chemical potential is discontinuous, this kind of
phase transition is called the second-order phase transition.
For Van der Waals equation, at the critical point no latent and sudden
change in volume happen between liquid phase and gas phase. According to
Ehrenfest classification, this kind of phase transition is continuous.

To understand the types of the phase transition for the GB-dS black hole, we should calculate the specific heat at constant pressure $C_P $, the
volume expansivity $\beta$, and the isothermal compressibility $\kappa $:

\begin{equation}
\label{eq35}
C_P =T_{eff} \left( {\frac{\partial S}{\partial T_{eff} }} \right)_{P_{eff}
}= {T_{eff}}{\left( {\frac{{\partial S}}{{\partial x}}\frac{{\partial x}}{{\partial {T_{eff}}}}} \right)_{{P_{eff}}}},
\ee
\be
\beta =\frac{1}{v}\left( {\frac{\partial v}{\partial T_{eff} }}
\right)_{P_{eff} }
=\frac{1}{v}\left( {\frac{\partial v}{\partial x}\frac{\partial x}{\partial
T_{eff} }} \right)_{P_{eff} },
\end{equation}
\be
\kappa _T =-\frac{1}{v}\left( {\frac{\partial v}{\partial P_{eff} }}
\right)_{T_{eff} }
=-\frac{1}{v}\left( {\frac{\partial v}{\partial x}\frac{\partial x}{\partial
P_{eff} }} \right)_{T_{eff} } .
\ee

\begin{figure}[htbp]
\center{\subfigure[~$ d=5,\tilde \alpha=0.1, P_c=0.008229$] {
\includegraphics[width=4cm,keepaspectratio]{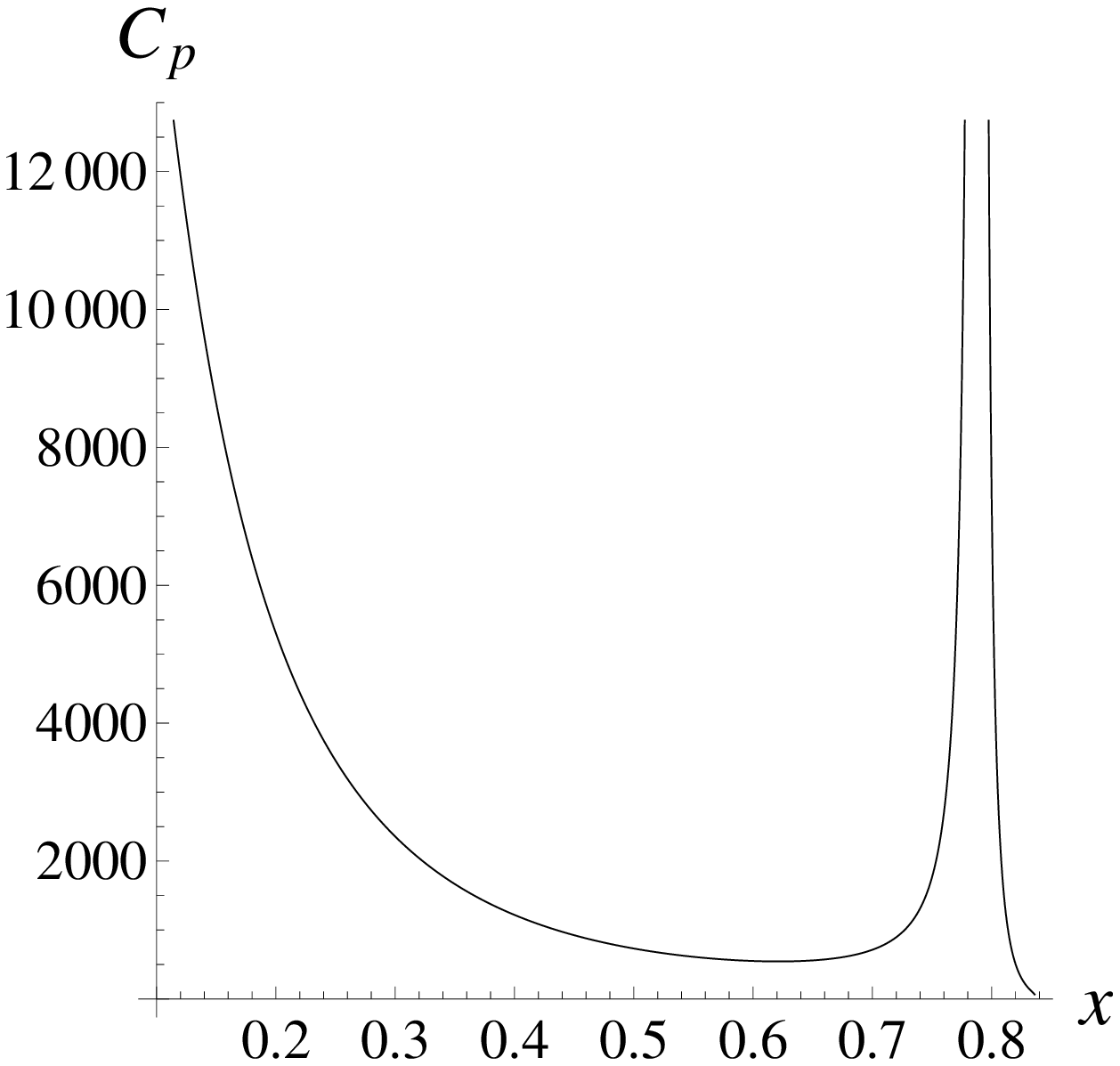}\hspace{0.8cm}}
\subfigure[~$ d=5 $, $\tilde \alpha=0.01$, $P_c=0.010356$] {
\includegraphics[width=4cm,keepaspectratio]{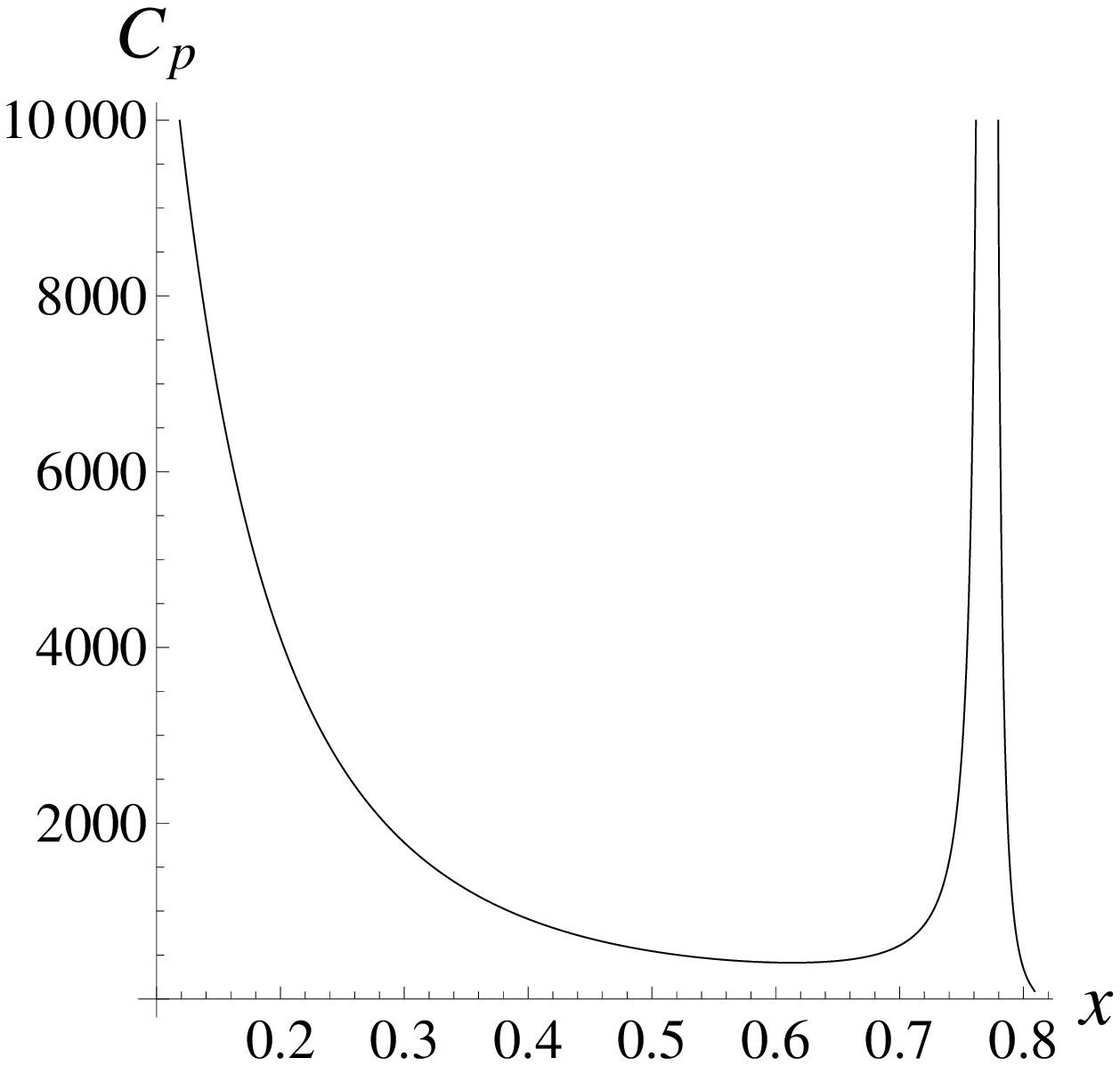}\hspace{0.8cm}}
\subfigure[~$ d=6, \tilde \alpha=0.1, P_c=0.031920$] {
\includegraphics[width=4cm,keepaspectratio]{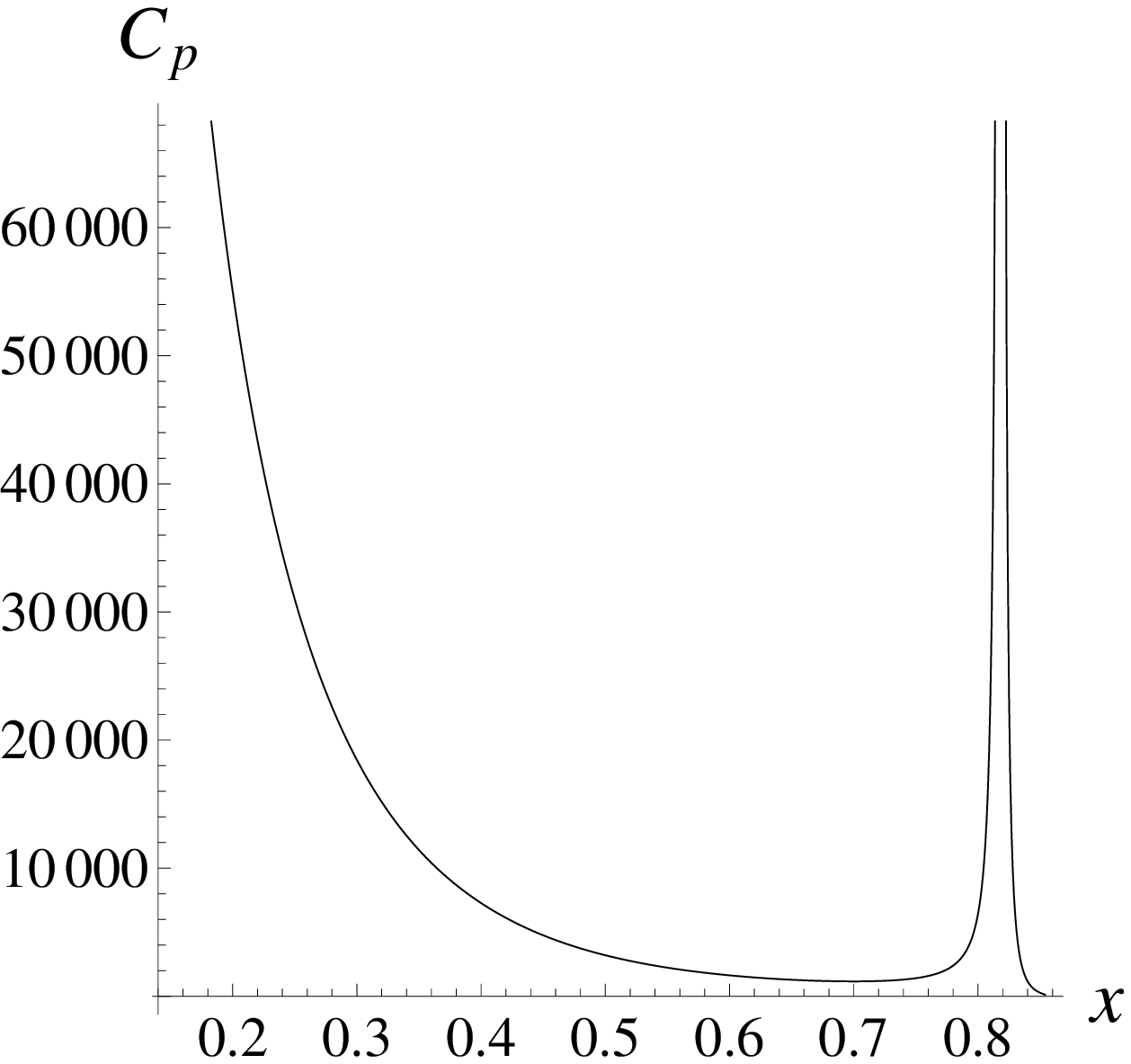}\hspace{0.8cm}}
\caption{$C_P-x$ curves for charged GB-dS black hole with $Q=1$.\label{cp}}}
\end{figure}

\begin{figure}[htbp]
\center{\subfigure[~$ d=5$, $\tilde \alpha=0.1$, $T_c=0.028628$] {
\includegraphics[width=4cm,keepaspectratio]{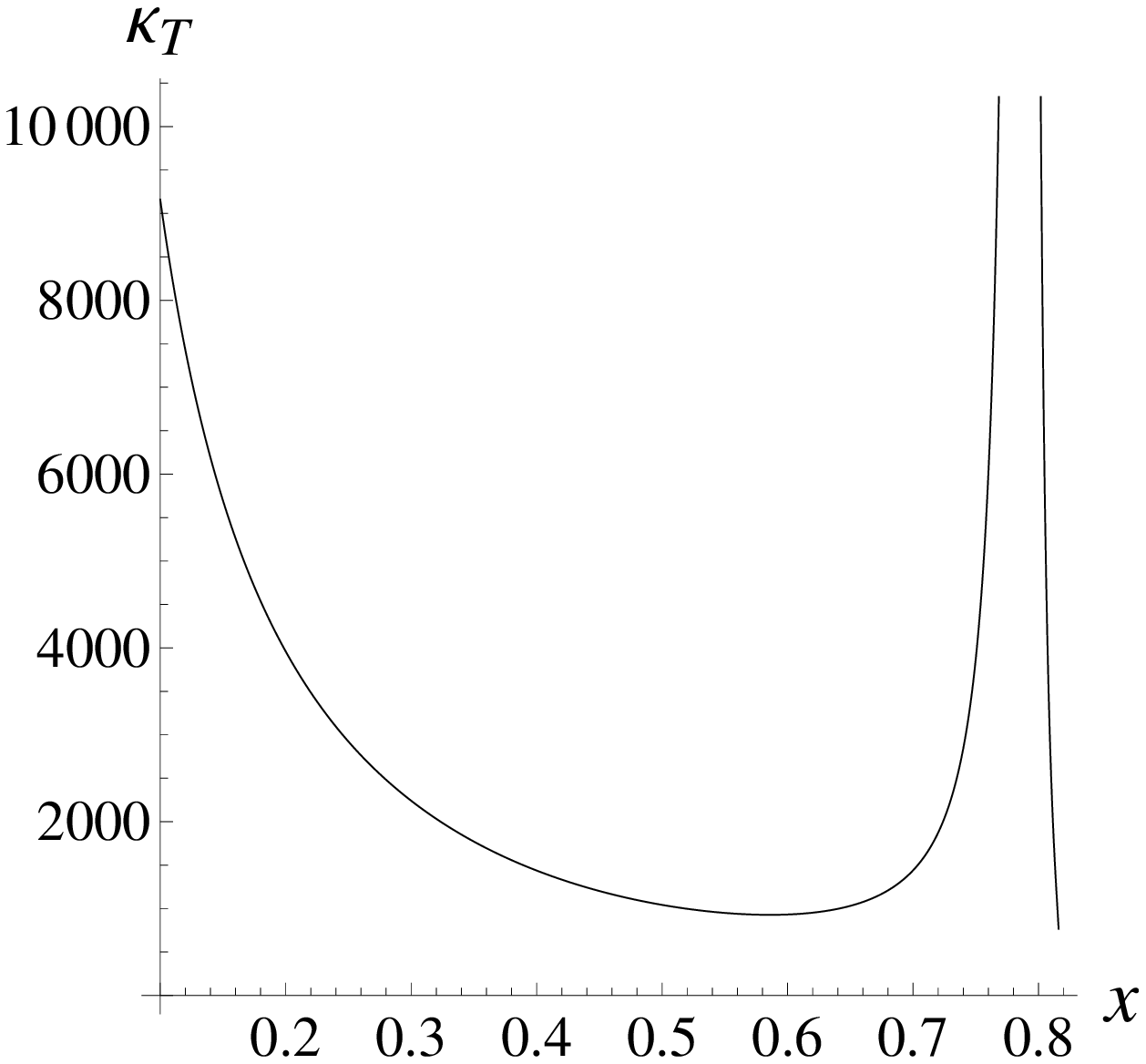}\hspace{0.8cm}}
\subfigure[~$ d=5 $, $\tilde \alpha=0.01$, $T_c=0.033651$] {
\includegraphics[width=4cm,keepaspectratio]{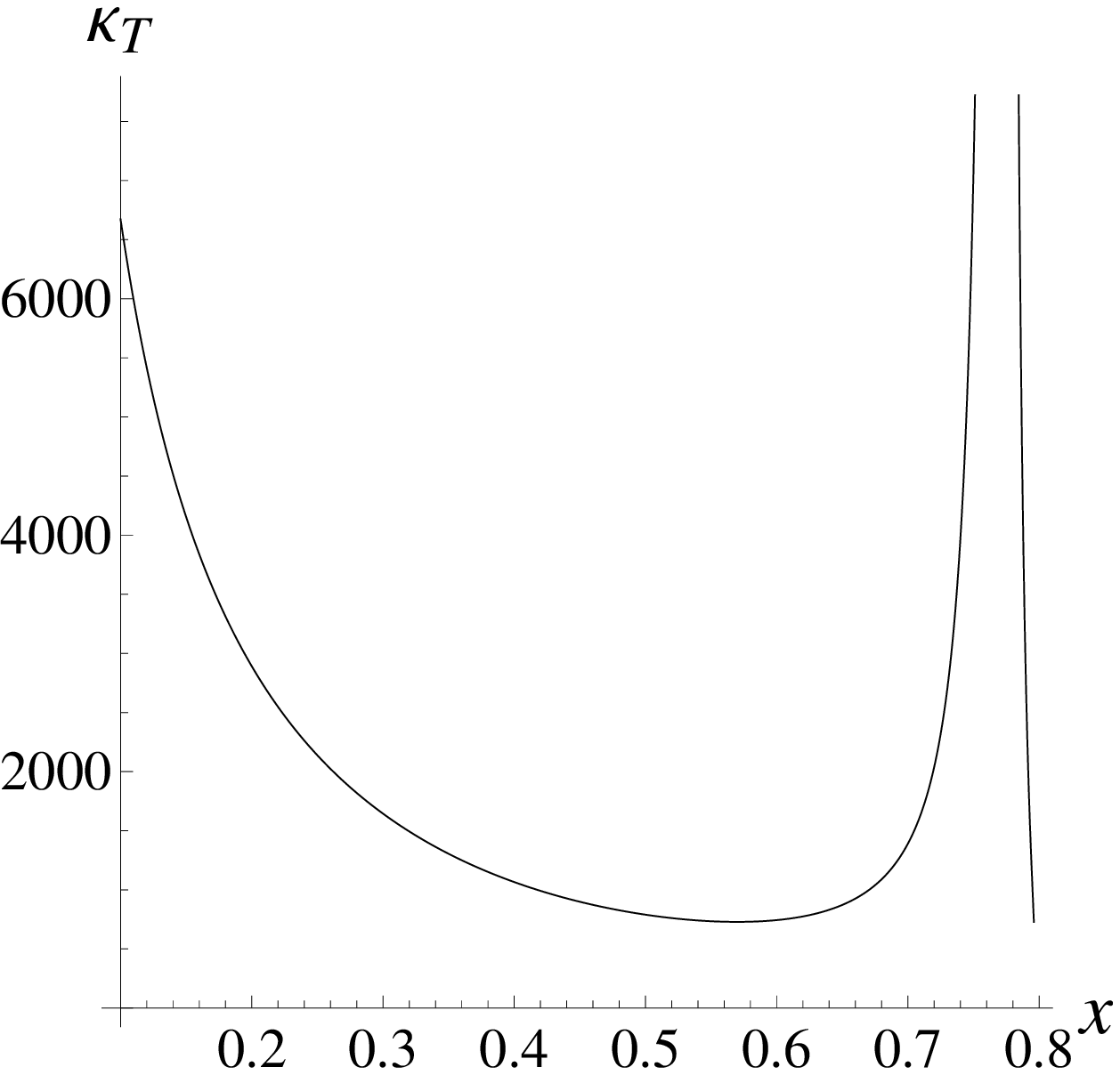}\hspace{0.8cm}}
\subfigure[~$ d=6$, $\tilde \alpha=0.1$, $T_c=0.040562$] {
\includegraphics[width=4cm,keepaspectratio]{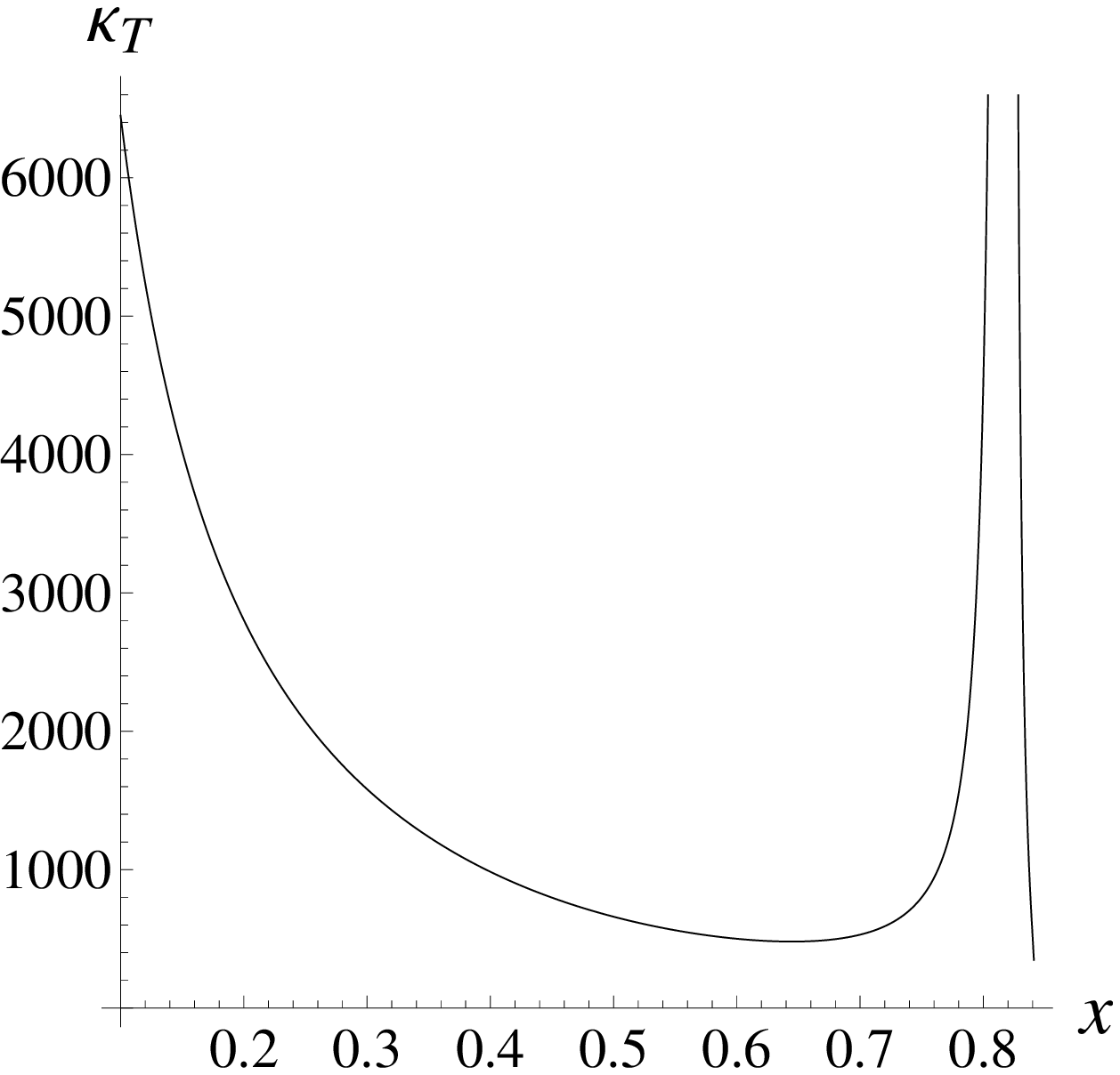}\hspace{0.8cm}}
\caption{$\kappa_T-x$ curves for charged GB-dS black hole  with $Q=1$.\label{kappa}}}
\end{figure}

\begin{figure}[htbp]
\center{\subfigure[~$ d=5 $, $\tilde \alpha=0.1$] {
\includegraphics[width=4cm,keepaspectratio]{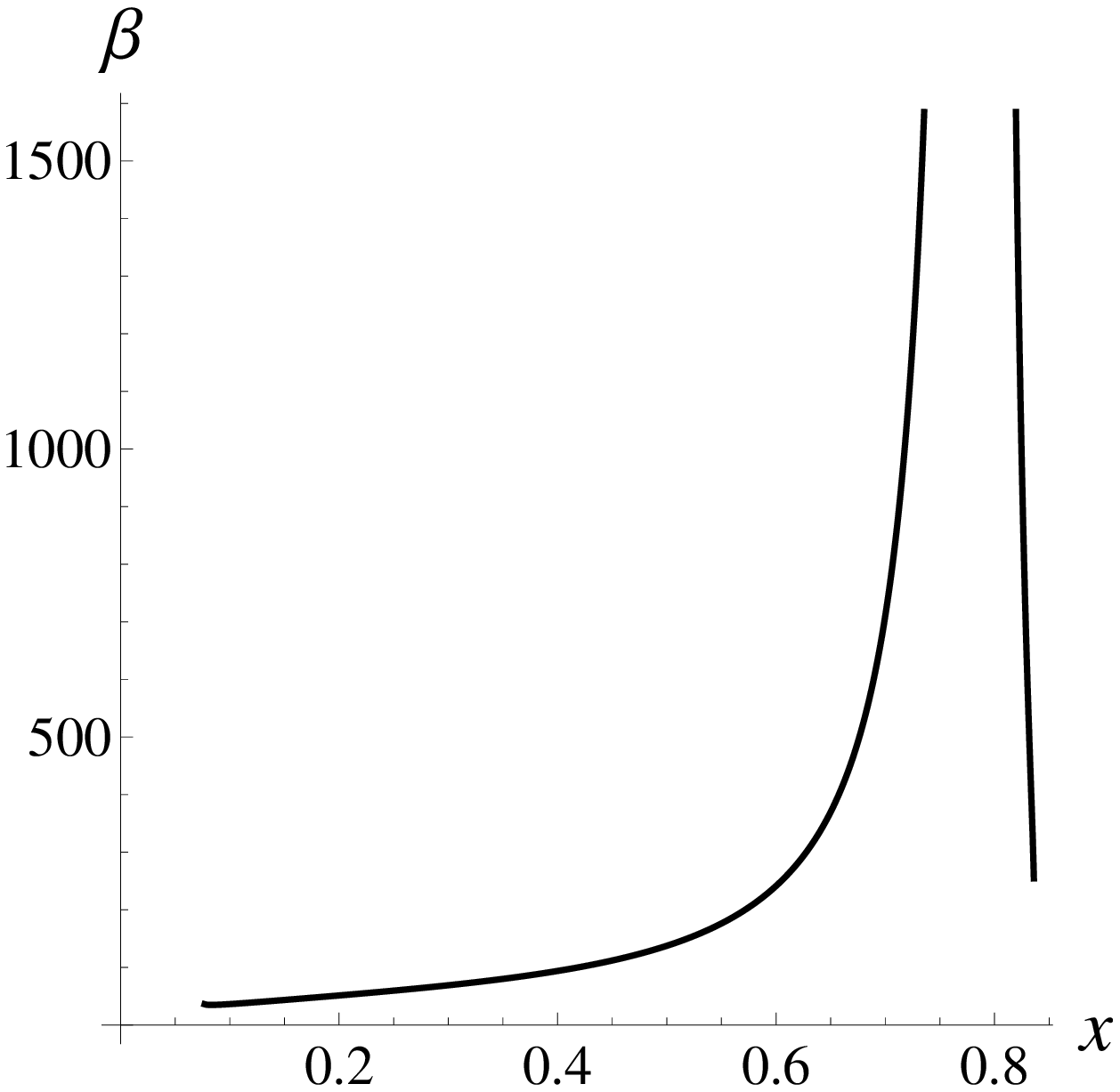}\hspace{0.8cm}}
\subfigure[~$ d=5 $, $\tilde \alpha=0.01$] {
\includegraphics[width=4cm,keepaspectratio]{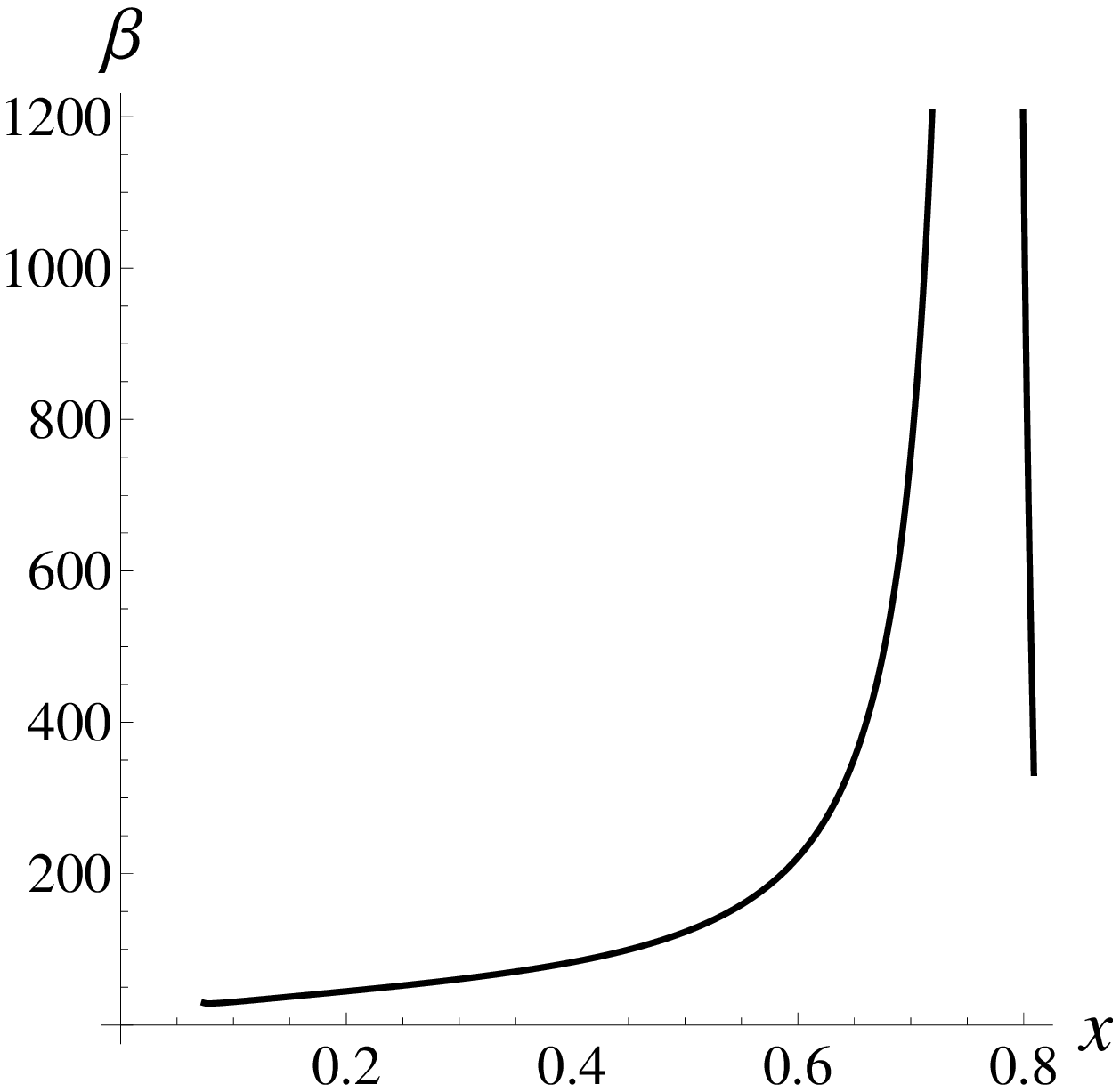}\hspace{0.8cm}}
\subfigure[~$ d=6 $, $\tilde \alpha=0.1$] {
\includegraphics[width=4cm,keepaspectratio]{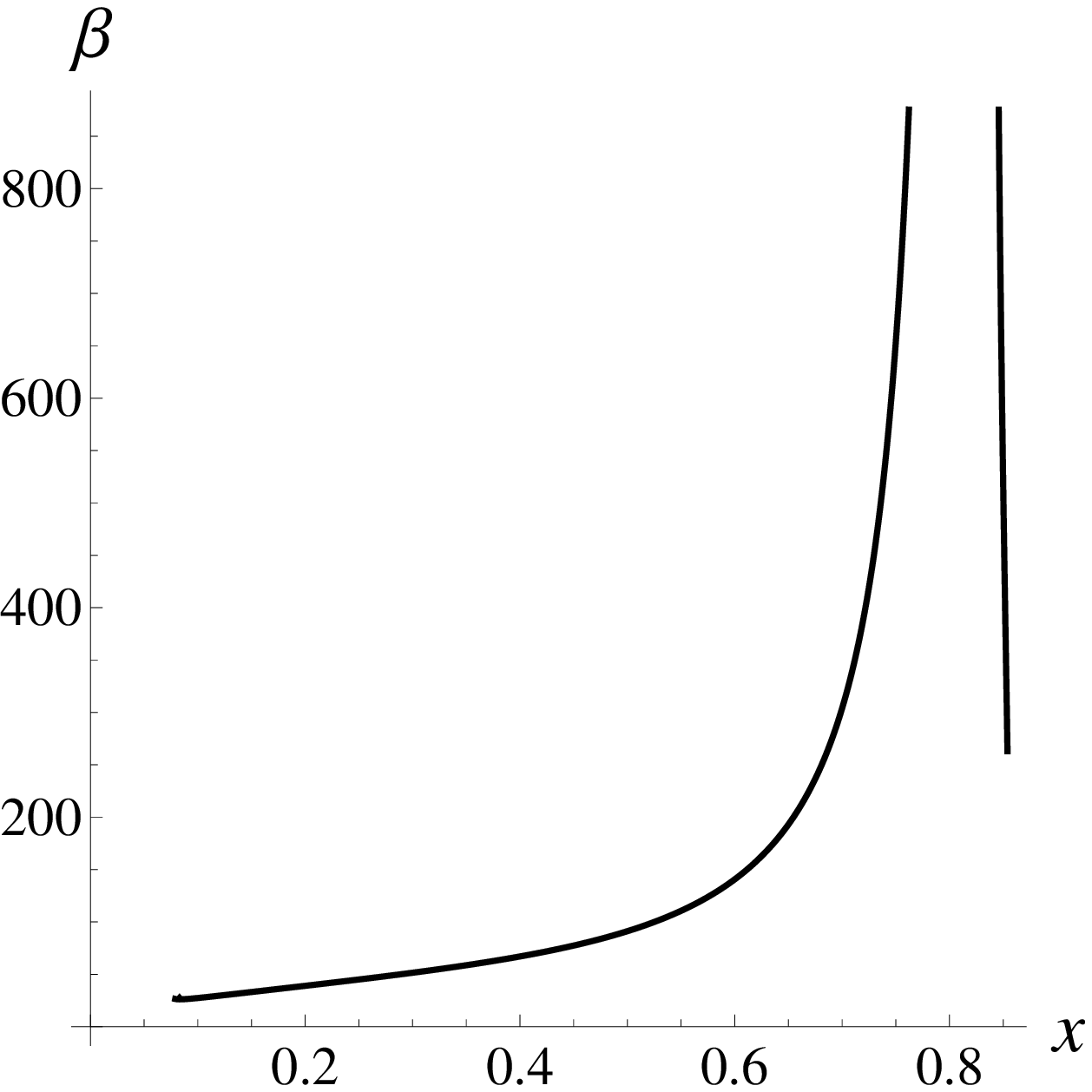}\hspace{0.8cm}}
\caption{$\beta-x$ curves for charged GB-dS black hole  with $Q=1$.\label{beta}}}
\end{figure}

\begin{figure}[htbp]
\center{\subfigure[~$ d=5 $, $\tilde \alpha=0.1$] {
\includegraphics[width=5cm,keepaspectratio]{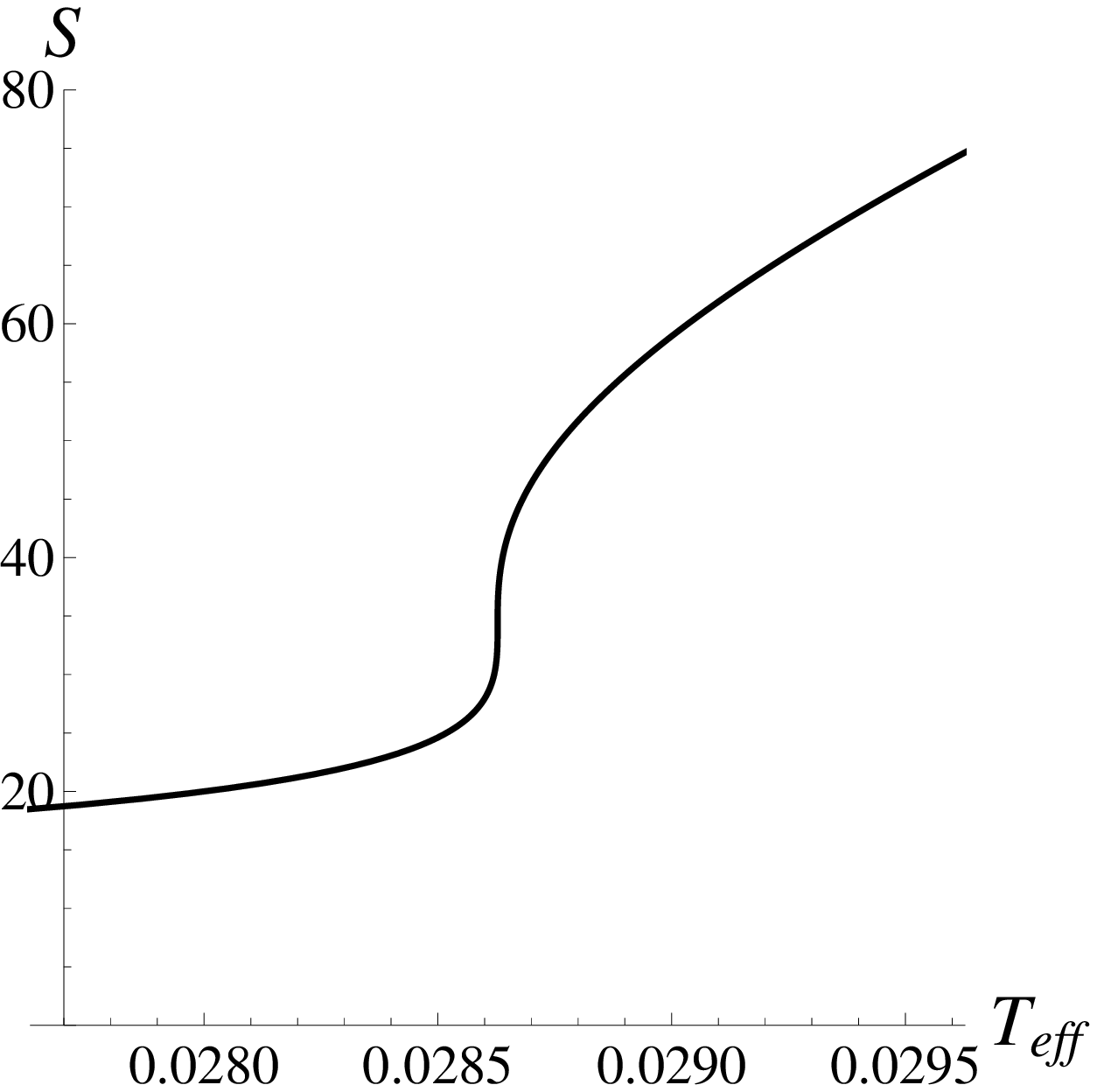}\hspace{0.8cm}}
\subfigure[~$ d=5 $, $\tilde \alpha=0.01$] {
\includegraphics[width=5cm,keepaspectratio]{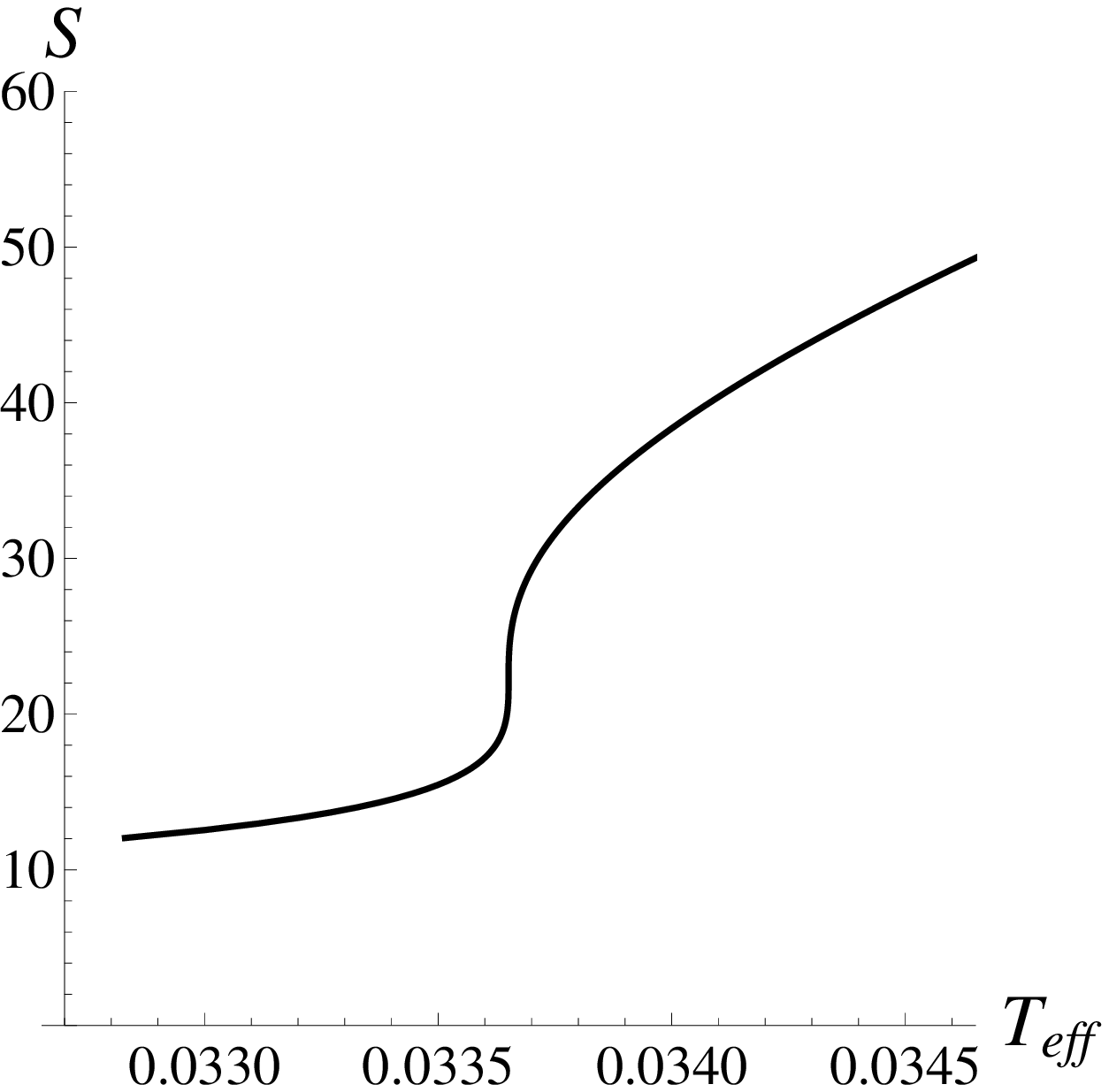}\hspace{0.8cm}}
\subfigure[~$ d=6 $, $\tilde \alpha=0.1$] {
\includegraphics[width=5cm,keepaspectratio]{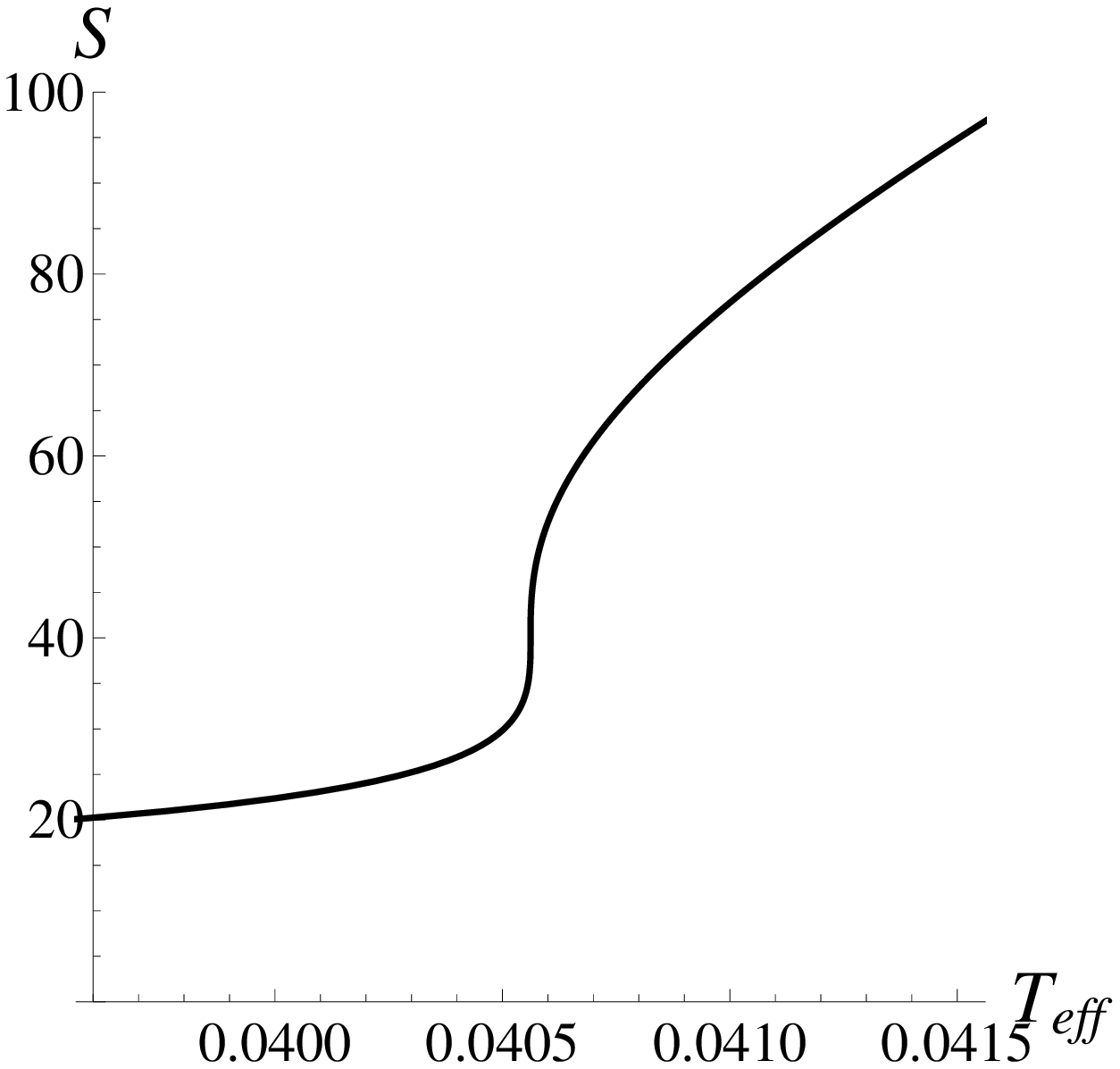}\hspace{0.8cm}}
\caption{$S-T_{eff}$ with $Q=1$.\label{st}}}
\end{figure}

We also depict the curves of $C_P-x$, $\kappa_T-x$ and $\beta-x$ in the Fig.\ref{cp}, Fig.\ref{kappa} and Fig.\ref{beta} respectively.
From these curves, we find that, for the charged GB-dS black hole, there exists an infinite peak in these curves. Moreover,
in Fig.\ref{st}, one can see that the curves of $S-T_{eff}$ are continuous at the critical point. According
to Ehrenfest, the phase transition of the charged GB-dS black
hole should be the second-order one, which is similar to the RN-(A)dS black hole.

\section{discussion and conclusion}

In this paper, we adopted Ehrenfest¡¯s classification to study the phase transition of the charged GB-dS
black hole.
After introducing the connection between the thermodynamic
quantities corresponding to the black hole horizon and the
cosmological horizon, we give the effective thermodynamic quantities, such as effective pressure $P_{eff}$,
effective temperature $T_{eff}$, etc, of the black hole. From the  equation of state described by the effective thermodynamic quantities, we study the $P-V$ criticality.
As illustrated in Fig.1, it exhibits a similar phase transition to Van der Waals equation.

By treating the effective cosmological constant as the thermodynamic pressure, in the
extended phase space, we completely follow the Ehrenfest classification to explore the type
of the phase transition of the charged GB-dS black hole. It is shown that when the entropy is a continuous function of temperature. However, the heat
capacity $C_P$, the isothermal compressibility $\kappa_T$ and the expansion coefficient $\beta$ are all divergent
at the critical point. This means this kind of phase transition for the charged GB-dS black hole  is second order.

\begin{acknowledgments}\vskip -4mm
This work is supported by NSFC under Grant Nos.(11475108,11175109;11075098;11205097)
and the doctoral Sustentation foundation of Shanxi Datong University
(2011-B-03).
\end{acknowledgments}

\end{document}